\documentclass[aps,prb,twocolumn,groupedaddress,bibnotes,citeautoscript]{revtex4-1}

\usepackage{graphicx}% Include figure files
\usepackage{dcolumn}% Align table columns on decimal point
\usepackage{bm}% bold math
\usepackage{color}

\begin{document}

%\preprint{}

\title{First-principles study of the mobility of SrTiO$_3$}
%\title{Effects of polar optical phonon scattering on the mobility of SrTiO$_3$}

\author{Burak Himmetoglu, Anderson Janotti, Hartwin Peelaers, Audrius Alkauskas, and Chris G. Van de Walle}
%\thanks{}
\affiliation{Materials Department, University of California, Santa Barbara, CA 93106-5050}

\date{\today}

\begin{abstract}

We investigate the electronic and vibrational spectra of SrTiO$_3$, as well as the coupling between them, using first-principles calculations.
We compute electron-phonon scattering rates
for the three lowest-energy conduction bands and use Boltzmann transport theory
to calculate the room-temperature mobility of SrTiO$_3$.
The results agree with experiment and highlight the strong impact of longitudinal optical phonon scattering.
Our analysis provides important insights into the key factors that determine
room temperature mobility, such as the number of conduction bands and the nature and frequencies of
longitudinal phonons. Such insights provide routes to engineering
materials with enhanced mobilities.
\end{abstract}

%\pacs{71.20.Ps,71.27.+a,71.38.Ht,61.72Bb}
%\keywords{}

\maketitle

%\section{Introduction}

There is great interest in using SrTiO$_3$ (STO) as a wide-band-gap semiconductor in novel electronic devices.
Thanks to recent progress in epitaxial growth of STO~\cite{stemmer-sto-mobility},
$n$-doped films have been achieved with carrier mobility as high as 53,000 cm$^2$V$^{-1}$s$^{-1}$ at $T$=2 K~\cite{stemmer-new}.
However, room-temperature
mobilities are orders of magnitude smaller, around a few cm$^2$V$^{-1}$s$^{-1}$,
potentially forming a significant limitation in electronic device applications~\cite{stemmer-sto-mobility}
and lending urgency to the investigation of the key material parameters that affect electron transport.
Electron mobility and other transport properties depend crucially on the conduction-band
structure of STO, which has already been investigated in a number of
studies~\cite{hwang-sto-fs,allen-sto-fs,stemmer-sto-band}.
Still, a clear understanding of the transport properties of STO has not yet been established.

%NEW ADDITION
Transport properties, conduction mechanisms, and the dependence of mobility on
temperature and electron density of STO have been investigated both in single crystal~\cite{Spinelli-STO}
and epitaxially grown samples~\cite{Jena-STO}.
%END NEW ADDITION
Recent experiments based on the measurement of Shubnikov de Haas oscillations~\cite{allen-sto-fs,hwang-sto-fs} in
STO thin films yielded effective band masses that are significantly larger than those obtained
from band-structure calculations~\cite{sto-hse-bs}. The mass enhancement was
attributed to the strong electron-phonon coupling in STO~\cite{allen-sto-fs,hwang-sto-fs}.
The effect of electron-phonon interactions has also been discussed in
relation to electron mobility.
% NEW ADDITION (sharp --> rapid)
Typically, the rapid decrease of the electron mobility with increasing temperature has
been associated with scattering of conduction electrons by polar optical phonon
modes~\cite{frederikse-sto,tufte-sto}.
Consequences of electron-phonon scattering on the effective mass of electrons in the conduction
bands have been studied by analyzing the spectral weight of the experimental optical
conductivity~\cite{mazin-sto-elph} and within the context of large-polaron models~\cite{devreese-sto-elph}.
%NEW ADDITION
More recently, the effect of electron-phonon scattering over a wide temperature range has been
investigated for thin films~\cite{Jena-STO}. An analysis based on phenomenological models is consistent with
longitudinal optical (LO) phonon scattering determining the room-temperature mobility,
while at lower temperatures (between 2 and 200 K), transverse optical (TO) modes were also found to be important.
%
%Using results from an effective mass approximation and
%model electron-phonon scattering rates, it has been shown that the low room temperature mobility
%can be attributed to scattering of electrons with longitudinal optical modes~\cite{Jena-STO}.
%Instead, between 2-200 K, an accurate fit to experimental data requires inclusion of scattering models
%that contain acoustic and transverse optical modes as well~\cite{Jena-STO}.
%While these findings provide some insight into the scattering mechanisms present in STO,
%they depend on multiple parameters that need to be fitted to the experimental data.
%In spite of all these studies of the mobility of STO,
%the impact of electron-phonon coupling on the transport properties has not yet been investigated based on first-principles calculations.
% END NEW ADDITION

Here we investigate the vibrational and electronic spectra of STO, as well as their coupling.
We find that polar optical mode scattering leads to low mobilities at room temperature.
%which sheds light on the experimental results in the literature.
%Interestingly, the scattering rates are quite different for the lowest three conduction bands in STO.
%The difference between scattering rates of different bands could be tuned,
Our analysis provide insights into the mechanisms that could potentially be used to tune the scattering,
for instance by nanostructuring, by using epitaxial strain
(which was already discussed in the context of effective masses~\cite{sto-hse-bs}),
or by replacing Ti with a heavier transition metal to increase spin-orbit splittings,
leading to possible enhancement of mobility.
%We find that the electrons occupying the heavy effective mass band scatter less effectively than
%electrons occupying the light effective mass bands. This results in a competition between effective masses and
%scattering rates to determine the mobility of electrons.
%This competition could be exploited using epitaxial strain, as was previously discussed in the context of effective masses~\cite{sto-hse-bs}, to design higher mobility interfaces.
%We suggest this can be exploited to enhance the mobility, for instance by nanostructuring or by using epitaxial strain (which was already discussed in the context of effective masses~\cite{sto-hse-bs}).

Very few first-principles calculations of mobilities including electron-phonon interactions
have appeared in the literature, for other materials \cite{restrepo,kaasbjerg-mos2,mbn-1}.
Obtaining converged scattering rates and transport integrals has been very challenging.
Our present work is based on accurate interpolation schemes and enabled by the use of an analytical
model for electron-phonon coupling, which we explicitly justify.
These developments allow us to obtain converged results at reasonable computational cost.

%\section{Computational Methods}

Our calculations are performed with the plane-wave self-consistent field (PWSCF) code
of the Quantum ESPRESSO package~\cite{QE}, using ultra-soft pseudopotentials~\cite{uspp}.
We use the local density approximation (LDA) with the Perdew-Zunger parametrization~\cite{pz};
we will verify that the use of the LDA provides adequate accuracy for the quantities of interest in this work.
Plane-wave basis sets are used to expand wavefunctions and charge densities, with kinetic energy cutoffs
of 50 Ry and 600 Ry, respectively. The phonon spectrum is calculated using density
functional perturbation theory (DFPT) as implemented in Quantum ESPRESSO~\cite{dfpt}.
Brillouin-zone (BZ) integrations are performed on a $8\times 8 \times 8$ special $k$-point
grid~\cite{mp} for self-consistent field calculations. The phonon spectrum is calculated on
a $4 \times 4 \times 4$ special point grid, and interpolated along lines connecting special
points within the BZ.
% in order to obtain the vibrational spectrum.
The splitting of longitudinal
and transverse optical modes at $\Gamma$ has been taken into account with the method of Born and Huang~\cite{born-huang}.
The Fermi integral for the calculation of the conductivity tensor is performed by
the BoltzTrap code~\cite{boltztrap}.

%\section{Phonon spectrum and band structure}

\begin{figure}[!ht]
\includegraphics[width=0.40\textwidth]{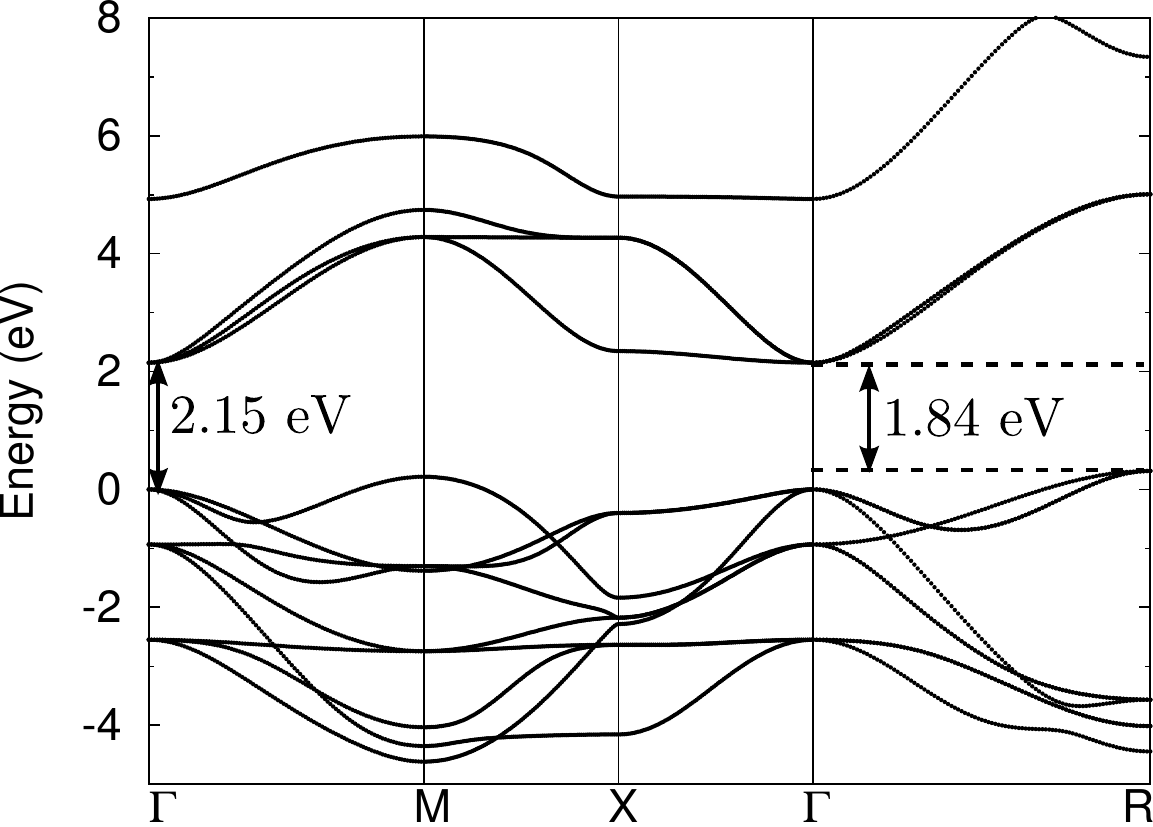}
\caption{\label{fig:band} Calculated band structure of cubic SrTiO$_3$ using the LDA functional.}
\end{figure}

We consider the cubic phase of SrTiO$_3$, which is the stable phase at room temperature. The optimized lattice parameter
is $a_0 = 3.86$ \AA, about $1\%$ smaller than experiment~\cite{cao-sto-lat} as expected from the LDA functional.
The LDA band structure is shown in Fig.~\ref{fig:band}.
The valence bands are derived from O $p$ orbitals, while the lowest three
conduction bands are derived from the Ti 3$d$ t$_{2g}$ orbitals ($d_{xy}$,
$d_{zx}$ and $d_{yz}$).
Due to the localized nature of these Ti 3$d$ orbitals, the conduction bands have low dispersion,
leading to high effective masses.
As expected, LDA underestimates the band gap, producing an indirect R-$\Gamma$ gap of $1.84$ eV and a direct $\Gamma$-$\Gamma$ gap of 2.15 eV,
compared to the experimental values of 3.25 eV and 3.75 eV, respectively~\cite{sto-gap}.  In all other respects, however,
LDA produces a reliable band structure.  In particular, the shape of the conduction bands and the effective masses are in good
agreement with more accurate calculations~\cite{sto-hse-bs} using the hybrid functional of Heyd, Scuseria and Ernzerhof (HSE) \cite{hse-1}, as can be seen from Table.~\ref{tab:bmass}.

\begin{table}[!ht]
\caption{\label{tab:bmass} Calculated band masses (in units of electron mass) along the
$\Gamma$-X and $\Gamma$-M directions, based on the LDA
and the HSE hybrid functional.  HSE values were taken from Ref.~\onlinecite{sto-hse-bs}.}
\begin{ruledtabular}
\begin{tabular}{ccccc}
%  & $(\Gamma-X)_{\rm LDA}$ & $(\Gamma-X)_{\rm HSE}$ & $(\Gamma-M)_{\rm LDA}$ & $(\Gamma-M)_{\rm HSE}$ \\
  & \multicolumn{2}{c}{$\Gamma$-X} & \multicolumn{2}{c}{$\Gamma$-M} \\
  & LDA & HSE & LDA & HSE \\
\hline
 $ m_1 $ & 6.11 & 6.10 & 0.77 & 0.85 \\
 $ m_2 $ & 0.35 & 0.39 & 0.56 & 0.64 \\
 $ m_3 $ & 0.35 & 0.39 & 0.35 & 0.39 \\
\end{tabular}
\end{ruledtabular}
\end{table}

\begin{figure}[!ht]
\includegraphics[width=0.40\textwidth]{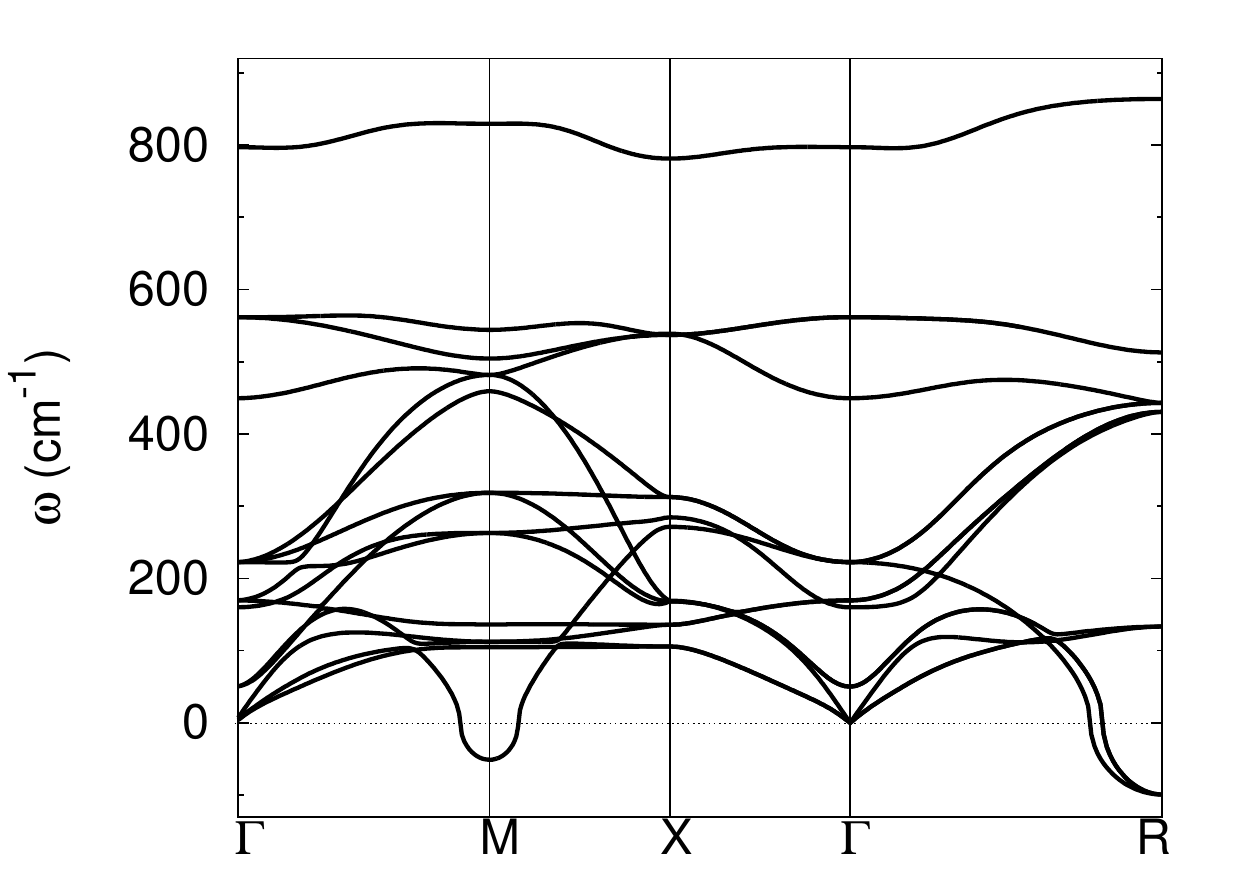}
\caption{\label{fig:phon} Calculated phonon spectrum of SrTiO$_3$ using the LDA functional.}
\end{figure}

The phonon spectrum obtained from DFPT is shown in Fig.~\ref{fig:phon}.
Table~\ref{tab:optph} lists the calculated optical phonon frequencies at $\Gamma$,
showing good agreement with experimental results.
The cubic phase of STO exhibits soft modes, as indicated by the presence of negative frequencies in the phonon spectrum (see Fig.~\ref{fig:phon}). These soft modes are related to structural instabilities associated with the distortions
that lead to the low-temperature tetragonal phase of STO.
The cubic phase is stabilized at finite temperature through strong anharmonic effects,
which are not included in our calculations based on the harmonic approximation.
%NEW ADDITION (at room temperature and for...)
These soft modes are not expected to be important at room temperature and for the types of electron-phonon scattering mechanisms
that are considered in the present study.
%
%Previous studies of the phonon spectrum in STO, using the Perdew-Burke-Ernzherof (PBE)
%functional~\cite{pbe}, yielded additional unstable modes at $\Gamma$~\cite{evarestov-sto-phon,lebedev-sto-phon}.
%This can be attributed to the fact that the PBE functional yields larger lattice parameters,
%thus lowering the phonon frequencies and
%leading to an off-center motion of the Ti atoms within the O octahedra, encoded as a soft mode at $\Gamma$.
%promoting a transition to a ferroelectric configuration~\cite{biegalski-sto-ferro,kim-sto-ferro}.
%The choice of the LDA functional in the present study results in stable phonons at $\Gamma$, and
%thus the room-temperature structure is intact for long-wavelength perturbations of the crystal.

%\begin{table}[!ht]
%\caption{\label{tab:optph} Calculated and experimental longitudinal optical (LO)
%and transverse optical (TO) phonon frequencies at the $\Gamma$ point,
%in units of cm$^{-1}$.}
%\begin{ruledtabular}
%\begin{tabular}{cccc}
%% $ \omega_{\rm LO}$ (LDA) & $ \omega_{\rm LO}$ (Exp.) & $ \omega_{\rm TO}$ (LDA) & $ \omega_{\rm TO}$ (Exp.) \\
%   \multicolumn{2}{c}{$ \omega_{\rm LO}$} & \multicolumn{2}{c}{$ \omega_{\rm TO}$} \\
%  LDA & Exp. & LDA & Exp. \\\hline
% 160 & 171 (Ref.~\onlinecite{sto-phon-1}) & 50 & 42 (Ref.~\onlinecite{sto-phon-1}), 91 (Ref.~\onlinecite{sto-phon-2}) \\
% 449 & 474 (Ref.~\onlinecite{sto-phon-1}) & 170 & 175 (Ref.~\onlinecite{sto-phon-1}), 170 (Ref.~\onlinecite{sto-phon-2}) \\
% 797 & 795 (Ref.~\onlinecite{sto-phon-1}) & 222 & 265 (Ref.~\onlinecite{sto-phon-1}) \\
%        &                             & 562 & 545 (Ref.~\onlinecite{sto-phon-1}), 547 (Ref.~\onlinecite{sto-phon-2})
%\end{tabular}
%\end{ruledtabular}
%\end{table}

\begin{table}[!ht]
\caption{\label{tab:optph} Calculated and experimental longitudinal optical (LO)
and transverse optical (TO) phonon frequencies at the $\Gamma$ point,
in units of cm$^{-1}$.}
\begin{ruledtabular}
\begin{tabular}{cccc}
% $ \omega_{\rm LO}$ (LDA) & $ \omega_{\rm LO}$ (Exp.) & $ \omega_{\rm TO}$ (LDA) & $ \omega_{\rm TO}$ (Exp.) \\
   \multicolumn{2}{c}{$ \omega_{\rm LO}$} & \multicolumn{2}{c}{$ \omega_{\rm TO}$} \\
  LDA & Exp. & LDA & Exp. \\\hline
 160 & 171$^a$ & 50  &  42$^a$, 91$^b$ \\
 449 & 474$^a$ & 170 & 175$^a$, 170$^b$ \\
 797 & 795$^a$ & 222 & 265$^a$ \\
        &      & 562 & 545$^a$, 547$^b$
\end{tabular}
\end{ruledtabular}
\begin{flushleft}
$^a$ Ref.~\onlinecite{sto-phon-1}.\\
$^b$ Ref.~\onlinecite{sto-phon-2}.\\
\end{flushleft}\end{table}

\begin{figure}[!ht]
\includegraphics[width=0.45\textwidth]{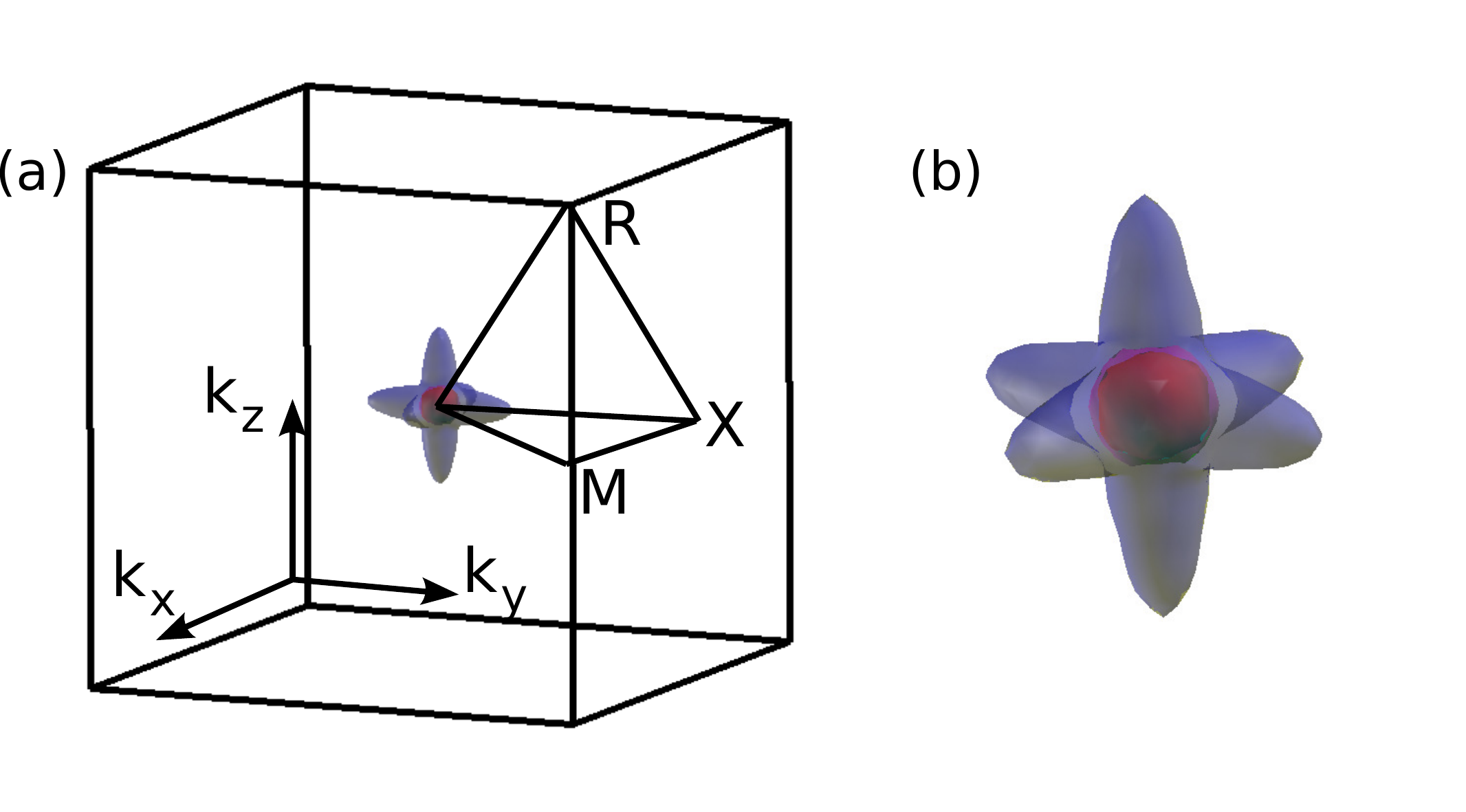}
\caption{\label{fig:fs}(color online) (a) Fermi surface of SrTiO$_3$ for an electron concentration $n=10^{20}$ cm$^{-3}$,
shown in the Brillouin zone, with high-symmetry points indicated. (b) Magnified view of the Fermi surface.
There are two low-mass bands with spherical shape (red) and one high-mass band with an ellipsoidal shape (blue).}
\end{figure}

Since the conduction-band minimum is located at $\Gamma$, the Fermi surface for $n$-doped STO is always
$\Gamma$-centered. Moreover, for experimentally achievable doping levels ($n < 10^{21}$ cm$^{-3}$),
the Fermi surface covers only a small portion of the BZ.
The smallness of the Fermi surface then requires a very fine sampling of the BZ.
In order to achieve the fine sampling, we have constructed maximally localized
Wannier functions~\cite{wan90} and interpolated the band structure of STO to a grid of $50 \times 50 \times 50$
$k$-points. The Fermi surface resulting from the Wannier interpolation for
%an electron doping level of
$n=10^{20}$ cm$^{-3}$ is shown in Fig.~\ref{fig:fs}.
%
%For electron doping levels encountered in experiments, the Fermi surface is centered around $\Gamma$, covering only a very small portion
%of the Brillouin Zone (BZ).
As a result, only phonons with very small wavevectors are able to scatter electrons.
In other words, only long-wavelength phonons contribute to the finite scattering time of the electrons.
Anharmonic effects that lead to stabilization of soft modes could introduce additional electron-phonon scattering
channels. These could be important around the temperature at which the tetragonal-to-cubic phase transition occurs (110 K, Ref. \onlinecite{cowley-sto}), but this is not the focus of our present study.
%NEW ADDITION
%While Ref.~\onlinecite{Jena-STO} considers scattering of electrons with TO modes, the model used for experimental
%fits does not contain effects of strong anharmonic effects. In fact, the impact of anharmonic interactions
%on electron-phonon scattering is not widely investigated in the literature and still poses some open questions.
% END NEW ADDITION

%\section{Electron-phonon interactions}

For polar crystals, electron-phonon scattering for phonons with small wavevectors ({\bf q})
is dominated by LO mode scattering~\cite{ziman}. It can be analytically shown that
LO mode scattering results in electron-phonon coupling matrix elements that have a $1/q$ dependence. Therefore,
for electrons in conduction bands near $\Gamma$ at temperatures where LO modes have large occupations,
this scattering channel dominates.
In this study we use the Fr\"ohlich model~\cite{ziman},
which provides a simple but accurate description of LO mode scattering with the
electron-phonon coupling matrix elements given by
%Numerically, the largest electron-phonon couplings for electrons near $\Gamma$
%are due to scattering of electrons with LO phonons. This can be
%seen from the Fr\"ohlich model that describe the coupling of electrons and LO
%phonons, and is given by~\cite{mahan}
%
\begin{equation}
g_{{\bf q} \nu} = \sqrt{ \frac{e^2\, \hbar \omega_{\nu}}{2 \epsilon_0\, V_{\rm cell}\, q^2}}\,
            \sqrt{ \frac{1}{\epsilon_{\infty}} - \frac{1}{\epsilon} }  \label{g_F}
\end{equation}
where $\omega_{\nu}$ are LO phonon frequencies, $\epsilon_0$ is the vacuum permittivity,
$V_{\rm cell}$ is the unit cell volume, and $\epsilon_{\infty}$ and $\epsilon$ are
the electronic and static dielectric constants, respectively.
%In this study, we consider the Fr\"ohlich model described by the above electron-phonon coupling
%matrix element to describe LO mode scattering.
In principle, longitudinal acoustic (LA) and TO modes also contribute to electron-phonon
scattering. However, for small wavevectors and at temperatures where LO mode occupancy is sizable, their
contribution is much smaller than that of LO modes~\cite{ziman}.
%However, for small wavevectors, their contribution is much smaller than that of LO modes~\cite{mahan}.
%For temperatures much smaller than LO mode frequencies (smaller than room temperature), LA and TO mode scattering can be important,
%but we are interested in room temperature mobilities, so we do not consider them in this study.
%While transverse optical and longitudinal acoustic modes could also lead to electron-phonon
%scattering, for small phonon wavevectors, their contribution is much smaller than that of LO modes.

Electrons in a state $\psi_{n{\bf k}}$ experience a scattering rate due to electron-phonon interactions
that can be obtained from Fermi's Golden rule as~\cite{ziman}:
\begin{eqnarray}
&& \tau_{n {\bf k}}^{-1} = \frac{2\pi}{\hbar}\, \sum_{{\bf q} \nu, m}\, \vert g_{{\bf q} \nu}({\bf k}, n, m) \vert^2\,
    \left( 1 - {\hat v}_{n {\bf k}} \cdot {\hat v}_{m {\bf k+q}} \right) \times  \nonumber\\
&& \quad \Big\{ \left( n_{{\bf q} \nu} + f_{m, {\bf k+q}} \right)\, \delta\left( \epsilon_{m, {\bf k+q}} - \epsilon_{n {\bf k}} - \hbar \omega_{{\bf q} \nu} \right)
  \nonumber\\
&& \quad + \left( 1 + n_{{\bf q} \nu} - f_{m, {\bf k+q}} \right)\, \delta\left( \epsilon_{m, {\bf k+q}} - \epsilon_{n {\bf k}} + \hbar \omega_{{\bf q} \nu} \right)
\Big\} \label{eqn:tau}
\end{eqnarray}
where $g_{{\bf q} \nu}({\bf k},n,m)$ are electron-phonon coupling matrix elements, the velocities ${\hat v}_{n {\bf k}}$ are defined below, and
$n_{{\bf q} \nu}$ and $f_{m, {\bf k+q}}$ are phonon and electron occupation factors described by Bose-Einstein and Fermi-Dirac
distributions respectively.
The delta functions ensure energy conservation.
%
%Due to the low dispersion of the conduction bands in STO,
%(i.e. bandwidth of Ti 3$d$-derived states being small),
%at electron doping levels within the experimental range
%($10^{18}$ cm$^3$ $< n <$ $10^{21}$ cm$^3$),
%the Fermi surface covers only a small portion of the BZ, as can be seen in Fig.~\ref{fig:fs}.
%Since the Fermi surface covers a small portion of the BZ (Fig.~\ref{fig:fs}),
Since the electrons all reside in the vicinity of the $\Gamma$ point (Fig.~\ref{fig:fs}),
only phonons with small wavevectors
can satisfy the energy conservation criterion between electronic states.
The Fr\"ohlich coupling given by Eq.~(\ref{g_F}) has no dependence on the electron wavevector ${\bf k}$;
ignoring the ${\bf k}$-dependence of the electron-phonon matrix element is a good approximation
since the Fermi surface occupies only a small portion of the BZ.
%In the case of Fr\"ohlich coupling given by Eq.~\ref{g_F}, the electron wavevector ${\bf k}$ is confined around
%$\Gamma$, so it corresponds to $g_{{\bf q} \nu}(\Gamma,n,m)$.
%Since the Fermi surface occupies a small region of BZ,
%ignoring the ${\bf k}$-dependence of electron-phonon couplings is a good approximation.
A further approximation inherent to the Fr\"ohlich model is the neglect of the dependence
of the electron-phonon coupling matrices on the band indices, resulting in
all scattering rates between the different conduction bands being equivalent.

The band velocities in Eq.~(\ref{eqn:tau}) are defined as
\begin{equation}
{\bf v}_{n {\bf k}} = \frac{1}{\hbar}\, \frac{\partial \epsilon_{n {\bf k}}}{\partial {\bf k}} \,\,\,\, , \,\,\,
{\hat v}_{n {\bf k}} = {\bf v}_{n {\bf k}} / \vert {\bf v}_{n {\bf k}} \vert \label{eqn:v}
\end{equation}
The velocity factor $(1 - {\hat v}_{n {\bf k}} \cdot {\hat v}_{m {\bf k+q}})$ in Eq.~(\ref{eqn:tau}) is added in an {\em ad hoc} manner~\cite{ziman} to include the effect of directionality in the transport scattering rate.
While the velocity factor has been shown to
be crucial in order to obtain the right temperature dependence of resistivity in simple metals~\cite{ziman},
we have explicitly checked that it does not produce significant modifications in our calculations for STO;
including the factor changes the scattering rate by only $\sim$10 \%.

%The smallness of the Fermi surface then requires a very fine sampling of the BZ in order to achieve convergence in
%integrating Eq.~\ref{eqn:tau}. In order to achieve the fine sampling, we have constructed maximally localized
%Wannier functions~\cite{wan90} and interpolated the band structure of STO into a grid of $50 \times 50 \times 50$
%$k$-points. The Fermi surface resulting from the Wannier interpolation for an electron doping level of $n=10^{20}$ cm$^{-3}$
%is shown in Fig.~\ref{fig:fs}.

The electron-phonon matrix elements in Eq.~(\ref{g_F}) can be evaluated using the calculated LO phonon mode frequencies
shown in Table~\ref{tab:optph} and the dielectric constants. The optical phonons have a rather flat spectrum (Fig.~\ref{fig:phon}),
therefore their dispersion can be ignored. Since $\epsilon \gg \epsilon_{\infty}$~\cite{dielectric-sto},
one can ignore the $1/\epsilon$ term.
The calculated dielectric constant $\epsilon_{\infty}=6.34$ is in reasonable agreement with the experimental
value of $\epsilon_{\infty} \simeq 5.59$~\cite{dielectric-sto}.
Evaluating the constants in Eq.~(\ref{g_F}), we can rewrite it as
\begin{equation}
g_{{\bf q} \nu} = \frac{C_{\nu}}{\left( q \cdot a_0/2 \pi \right)} \label{g_F2}
\end{equation}
The constants $C_{\nu}$ are calculated as $C_1 = 0.0024\, {\rm eV}^2$,
$C_2 = 0.0066\, {\rm eV}^2$, and $C_3 = 0.012\, {\rm eV}^2$,
where $C_{\nu}$ are indexed in order of increasing frequency, as listed in Table~\ref{tab:optph}.

%
%\section{Electron scattering time and mobility}

\begin{figure}[!ht]
\includegraphics[width=0.40\textwidth]{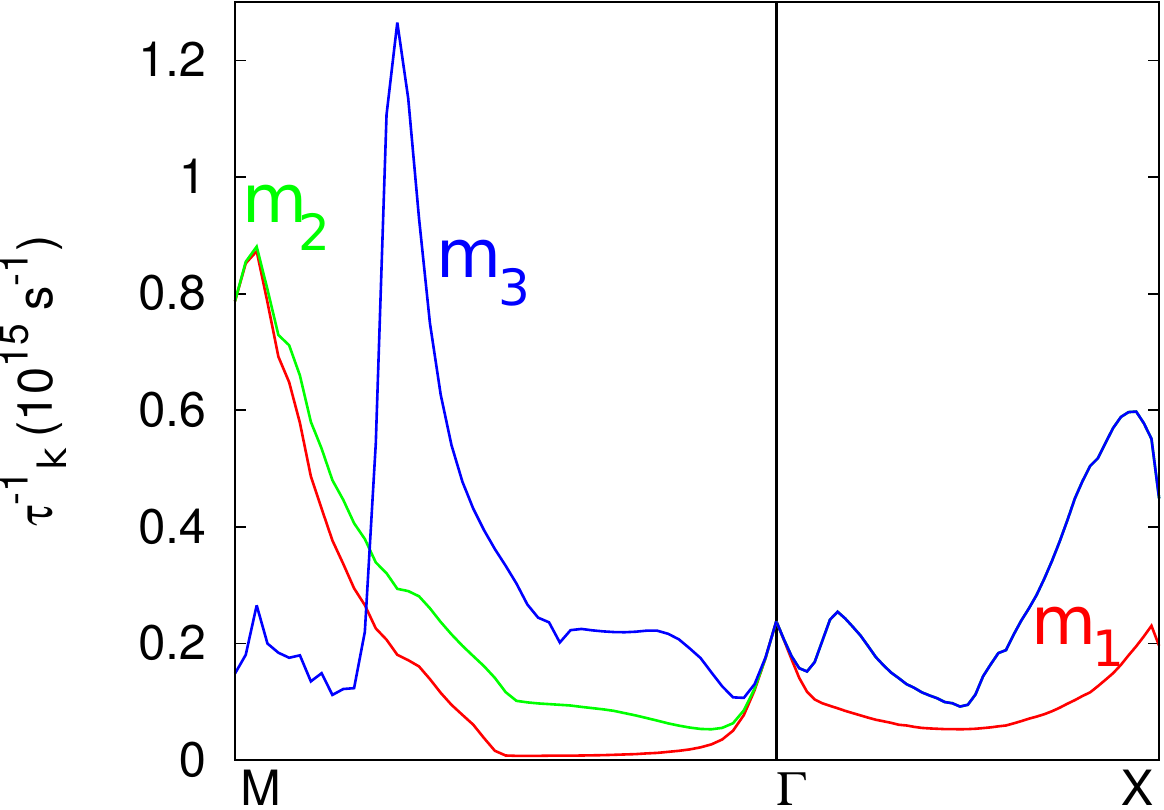}
\caption{\label{fig:tau}(color online) Electron-phonon scattering rates for electrons with a given momentum between
$\Gamma$ and M or between $\Gamma$ and X, for the three conduction bands, at 300 K and for $n=10^{20}$cm$^{-3}$.
\label{fig:scatt}}
\end{figure}

Using the electron-phonon coupling matrices of Eq.~(\ref{g_F2}),
we obtain the electron-phonon scattering rates from Eq.~(\ref{eqn:tau}) for the three conduction bands
with effective masses listed in Table~\ref{tab:bmass}. We performed the calculations at 300 K,
using a $50\times50\times50$ grid of $q$-points. The energy-conserving
delta functions are replaced by Gaussians with a width of $0.05$ eV.
The choice of the $q$-point grid and smearing yields an accuracy of 0.02 fs
for the calculated scattering times. The resulting scattering rates are shown in Fig.~\ref{fig:tau}
along the high-symmetry directions $\Gamma$-X and $\Gamma$-M.

\begin{figure}[!ht]
\includegraphics[width=0.35\textwidth]{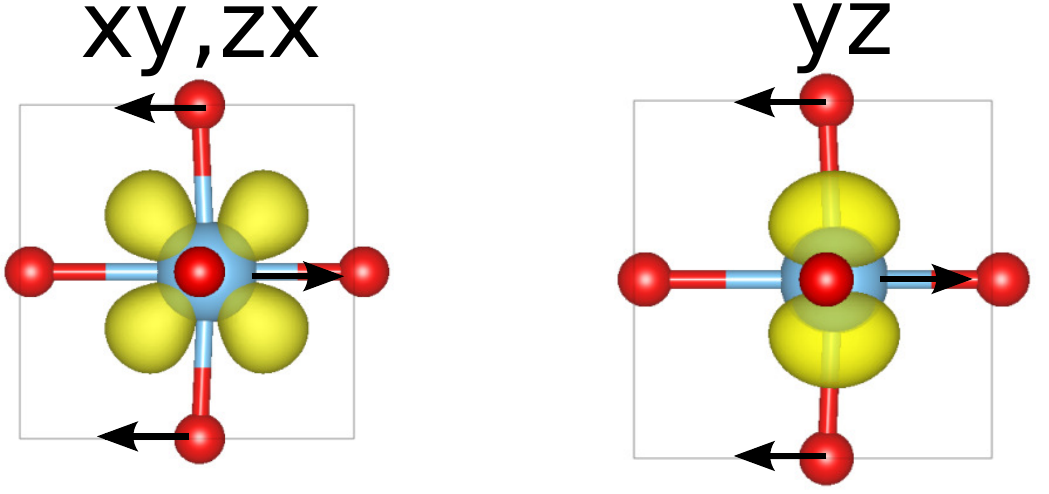}
\caption{\label{fig:kpdos}(color online)
Atomic displacements corresponding to the highest frequency LO mode
and t$_{2g}$ orbital isosurfaces representing the conduction bands.}
\end{figure}

Electrons with heavy masses (mass $m_1$) have the lowest scattering rate in most of the BZ, while lighter
electrons are scattered more effectively. This result can be understood by investigating
the effect of atomic displacements associated with polar optical modes on the band structure.
We show in Fig.~\ref{fig:kpdos} the displacements of atoms for the
highest frequency polar optical mode ($\omega_1 = 797$ cm$^{-1}$). The
relative displacements of the atoms are obtained from the eigenvectors
of the dynamical matrix at $\Gamma$.
%The heaviest conduction band is derived from
%the $d_{yz}$ orbital of Ti, while the $d_{xy}$- and $d_{zx}$-derived bands are lighter.
%Notice that the heavy mass is picked up by the choice of the LO phonon mode. All directions
%are equivalent due to cubic symmetry, when the LO mode is chosen as shown in Fig.~\ref{fig:kpdos},
%the symmetry is broken and the heavy band becomes the $d_{yz}$ orbital derived one.
In cubic symmetry, the $d_{xy}$, and $d_{yz}$, and $d_{zx}$ orbitals of Ti are in principle equivalent.
But when a particular LO mode is chosen. e.g., pointing along $x$, as in Fig.~\ref{fig:kpdos},
the symmetry is broken and the heavy band becomes the one derived from the $d_{yz}$ orbital,
while the $d_{xy}$- and $d_{zx}$-derived bands are lighter.
The O-atom displacements corresponding to this mode bring the O $p$ orbitals closer
to $d_{xy}$ and $d_{zx}$, resulting in strong
repulsion between filled O $p$ shells and conduction electrons occupying these two
t$_{2g}$ bands. The strong repulsion results in
a scattering rate which is larger for the low-mass band (Fig.~\ref{fig:tau}). On the other hand,
O $p$ and $d_{yz}$ orbitals do not approach each other during displacements
arising from the polar optical mode, and thus the scattering rate is smaller for the
heavy band (Fig.~\ref{fig:tau}).

The actual mobility of electrons is determined both by the scattering rates
and the effective masses. In order to obtain the room-temperature mobility,
we compute the conductivity tensor within Boltzmann transport theory given by~\cite{ziman}
\begin{equation}
\sigma_{\alpha\, \beta} = \frac{2\, e^2}{V_{\rm cell}}\, \sum_{n, {\bf k}}\,
  \tau_{n {\bf k}}\, \left( - \frac{\partial f_{n {\bf k}}}{\partial \epsilon_{n {\bf k}}} \right)\,
  v_{n {\bf k}, \alpha}\, v_{n {\bf k}, \beta} \label{eqn:cond}
\end{equation}
where $\alpha$ and $\beta$ are Cartesian indices.
Since STO is cubic, the conductivity tensor is diagonal with equal entries.
To evaluate Eq.~(\ref{eqn:cond}), we assume
the scattering time $\tau_{n {\bf k}}$ is given by its value at $\Gamma$.
% NEW ADDITION
This approximation is again justified by the smallness of
the region of the BZ occupied by the Fermi surface.
We compute the scattering rates and electron mobilities using $\mu=\sigma / ne$  at 300 K for a range of carrier densities between $10^{18}$ cm$^{-3}$ and $10^{21}$ cm$^{-3}$.
%, as reported in Table~\ref{tab:mu}.
We find mobilities between 7 and 18 cm$^2$/V/s,
in good agreement with experimental values~\cite{stemmer-sto-mobility,stemmer-new,Spinelli-STO,frederikse-sto,tufte-sto}.
%
%\begin{table}[!ht]
%\caption{\label{tab:mu} Calculated scattering rates at $\Gamma$ and electron mobilities at
%300 K for a range of carrier densities $n$.}
%\begin{ruledtabular}
%\begin{tabular}{cccc}
% n (cm$^{-3}$) & $\tau_{\Gamma}$ (fs) & $\mu$ (cm$^{2}$/V/s) \\
%\hline
%   $10^{18}$  &  6.46 & 17.8 \\
%   $10^{19}$  &  6.17 & 17.4 \\
%   $10^{20}$  &  4.22 & 13.1 \\
%   $10^{21}$  &  1.79 & 7.9
%\end{tabular}
%\end{ruledtabular}
%\end{table}
% END NEW ADDITION

%
%For 300 K and $n=10^{20}$cm$^{-3}$, using $\mu=\sigma / ne$,
%we find an electron mobility of 13.2 cm$^2$/V/s.
%This is in very good agreement with the experimental values, which are
%around a few cm$^{2}$ V$^{-1}$ s$^{-1}$~\cite{stemmer-sto-mobility,frederikse-sto,tufte-sto}.

%\begin{figure}[!ht]
%\includegraphics[width=0.40\textwidth]{SrTiO3_Sig.pdf}
%\caption{\label{fig:sig} Calculated electron conductivity as a function of Fermi energy.
%The zero of Fermi energy is set to the conduction-band minimum.
%The dashed lines correspond to the Fermi energy at 300 K and $n=10^{20}$cm$^{-3}$. }
%\end{figure}
%
%\section{Conclusions}

The insights provided by our calculations allow us to suggest several routes to achieving enhanced mobilities
by engineering materials or using alternative compounds.

\noindent
(1) Equation~(\ref{eqn:tau}) indicates that electron-phonon
scattering is ineffective when the LO phonon energy does not match the
energy difference between the initial ($\epsilon_{n {\bf k}}$)
and the final band energies ($\epsilon_{m, {\bf k+q}}$).
% due to phonon emission or absorption (which can occur via an inter-band or an intra-band scattering process).
Splitting the bands by an amount that exceeds the
LO phonon frequencies will reduce the scattering rates and result in
an increase in electron conductivity. This is due to the fact that
the number of available final states where an electron is scattered by a phonon
becomes smaller.
Given that the maximum LO phonon frequency
is around 0.1 eV, such splitting can be achieved by replacing Ti with a heavier
transition metal that would yield a splitting due to stronger spin-orbit coupling.

\noindent
(2) Nanostructuring can be used to exploit quantum confinement and
discretize the electronic spectrum with energy differences that do not match
phonon frequencies. Similar physics has been discussed in the context of quantum dots,
where it is known as the phonon-bottleneck effect\cite{phbott}.

\noindent
(3) The scattering rate can be reduced by decreasing the number of (nearly) degenerate
conduction bands. Perovskites with a conduction band derived from a single orbital,
such as BaSnO$_3$, display much higher room-temperature mobilities~\cite{bso}.
Part of this effect is of course due to the fact that in BaSnO$_3$ the conduction band is
derived from $s$ states, providing a smaller effective
electron mass; however, the reduction in the number of scattering channels for electrons via
optical phonons also contributes significantly to the enhancement in mobility.
In order to demonstrate this effect, we have evaluated the scattering rates
in Eq.~(\ref{eqn:tau}) using only a single band (with mass $m_1$). We find
that the reduction of the number of conduction bands from 3 to 1 (assuming
a fixed Fermi-level position) yields $\tau^{-1}_{\rm 3 bands}/\tau^{-1}_{\rm 1 band} \simeq 2.05$
at $\Gamma$.

\noindent
(4) Strain can be used to modify the shape and degeneracy of conduction bands.
% NEW ADDITION
In fact, a factor of 2 to 3 increase in mobility was observed under compressive strain in STO at low temperatures~\cite{stemmer-sto-band}.
Strain modifies the effective masses~\cite{sto-hse-bs}, which by itself can lead to
changes in mobility.
In addition, strain can break the degeneracy of the
three lowest energy conduction bands, which scatter differently.
% END NEW ADDITION
The strain that can be achieved by pseudomorphic growth is probably not large enough to split
the conduction bands by an amount larger than the LO phonon frequency~\cite{sto-hse-bs},
but it can still lead to modest enhancements in mobility.
The three lowest energy conduction bands in STO scatter differently
with LO phonons, yielding different scattering rates along the BZ (see Fig.~\ref{fig:scatt}).
Biaxial strain breaks the degeneracy of the bands at $\Gamma$ and will
modify the scattering rates (which in the unstrained case are very similar for the different bands, close to $\Gamma$).

\noindent
(5) Strain also modifies the LO phonon frequencies.
As can be seen from Eqs.~(\ref{g_F}) and (\ref{eqn:tau}), the scattering rate of electrons from LO phonons
is proportional to the phonon frequency. Tensile strain along the atomic
displacements of the LO modes (see Fig.~\ref{fig:kpdos}) could
lower the force constants and lead to lower frequencies. In addition,
lower LO phonon frequencies make it harder for electrons to reach the next
available conduction band, which acts as an additional
mechanism to reduce scattering.
Pseudomorphic growth on specific substrates \cite{sto-hse-bs} could thus be a powerful tool for engineering mobilities.
%Combined with the findings of Ref.~\onlinecite{sto-hse-bs}, it could be
%possible to propose substrates for STO growth that would lead
%to enhanced mobilities.

%Our work also proposes a simple methodology for calculating room temperature
%mobilities of polar materials. Calculation of mobilities from first-principles
%including electron-phonon interactions
%have been a serious challenge, and only a few examples exist in the literature
%with limited success~\cite{restrepo,kaasbjerg-mos2,mbn-1,mbn-2}. Due to the requirement of fine meshes
%for the convergence of scattering rates (Eq.~\ref{eqn:tau}) and transport integrals (Eq.~\ref{eqn:cond}),
%calculations have always been hampered by heavy computational load. Instead,
%we rely on accurate interpolation schemes~\cite{wan90,boltztrap} and an analytical
%model for electron-phonon coupling matrices (Eq.~\ref{g_F}), which overcome
%convergence problems without significant computational load.
%Improvements such as inclusion of the full $k$-dependent scattering rates $\tau_{n {\bf k}}^{-1}$
%in the transport integrals are rather straightforward and could lead to further
%predictions, such as full temperature dependence of mobility.

In conclusion, we have developed a computationally tractable first-principles methodology for calculating
mobilities of polar materials. We have explicitly
shown that the room-temperature mobility of electrons in STO is determined
by longitudinal optical mode scattering, and provided insights into avenues for
improving mobility.
%Using first-principles calculations, we have shown that the room temperature electron mobility
%of STO is limited by strong longitudinal optical phonon scattering, as deduced from
%experimental studies. We have investigated the effect of longitudinal optical
%phonon scattering on electrons that occupy the three conduction bands.
%We found that while heavy conduction
%electrons scatter less (have a longer scattering time) than the light electrons, their large effective
%mass acts as a bottleneck for room temperature mobility.
%
%Inclusion of strain the reduce the effective masses, and possibly tuning of LO mode frequencies could
%lead to enhanced mobilities for STO thin films, which could have profound consequences
%for electronic device applications.

%\section{Acknowledgements}
This work was supported by the MURI program of the Office of Naval Research, Grant Number N00014-12-1-0976.
Additional support was provided by ARO (W911-NF-11-1-0232).
Computational resources were provided by the Center for Scientific Computing at the CNSI and MRL (an NSF MRSEC, DMR-1121053) (NSF CNS-0960316), and by the Extreme Science and Engineering Discovery Environment (XSEDE), supported by NSF (OCI-1053575 and DMR07-0072N).

%
% HERE, mention anharmonicity, spin-orbit introducing opt-phonon scattering channels, umklapp

\bibliography{hjpavdw}

\end{document}